\documentclass[twocolumn,apl,reprint,showpacs,preprintnumbers,amsmath,amssymb,superscriptaddress,showkeys]{revtex4-2}
\usepackage{graphicx}
\usepackage{dcolumn}
\usepackage{bm}
\usepackage{amsmath}
\usepackage{amssymb}
\usepackage{braket}
\usepackage{todonotes}
\usepackage{color}
\usepackage{bbm}
\usepackage{mathtools}
\usepackage{epsfig}
\usepackage{epstopdf}
\usepackage{color}
\usepackage[english]{babel}
\usepackage{bm}
\usepackage{subfigure}
\usepackage{hyperref}
\usepackage{xcolor}
\usepackage{bbold}

\usepackage[autostyle]{csquotes}

\usepackage[normalem]{ulem}
\hypersetup{
    colorlinks,
    citecolor=blue,
    filecolor=black,
    linkcolor=blue,
    urlcolor=blue
}
 \usepackage{relsize}

\newcommand*{\rom}[1]{\expandafter\@slowromancap\romannumeral #1@}
\begin{document}

\title{Bloch dynamics in inversion symmetry broken monolayer phosphorene}
\author{Abdullah Yar*}
\affiliation{%
Department of Physics, Kohat University of Science and Technology,\\
Kohat-26000, Khyber Pakhtunkhwa, Pakistan\\
}%
\email{abdullahyardawar@gmail.com}
\author{Rifat Sultana}
\affiliation{%
Department of Physics, Kohat University of Science and Technology,\\
Kohat-26000, Khyber Pakhtunkhwa, Pakistan\\
}%

\date{\today}
\begin{abstract}
We investigate Bloch oscillations of wave packets in monolayer
phosphorene with broken inversion symmetry.
We find that the real space trajectories, Berry and
group velocities of Bloch electron undergo Bloch
oscillations in the system. The strong dependence
of Bloch dynamics on the crystal momentum is illustrated.
It is shown that the spin-orbit interaction crucially affects
the dynamics of the Bloch electron.
We also demonstrate the dynamics in external electric
and magnetic field within the framework of Newton's equations of motion, leading
to the geometric visualization of such an oscillatory motion.
In the presence of both applied in-plane electric and transverse magnetic fields,
the system undergoes a dynamical transition from confined
to de-confined state and vice versa, tuned by
the relative strength of the fields.
\end{abstract}

\maketitle

\section{Introduction}\label{Introduction}
Phosphorene is realized as an allotropic form of a monolayer black
phosphorus (BP) that has been the focus of intensive research efforts.
Its exotic electronic properties arise due to its highly anisotropic
nature originating from its puckered lattice structure~\cite{Rodin-PRL.112:176801,Low-PRL.113:106802,Qiao-NC.5:4475,Xia-NC.5:4458,Wang-NN.10:517}.
It belongs to the $D^{18}_{2h}$ point group, which has reduced symmetry
compared with its group IV counterparts having the $D^4_{6h}$ point group symmetry.
This class of quantum matter provides a unique platform to
study the fundamental many-body
interaction effects, high charge carrier mobility and exotic
anisotropic in-plane electronic properties.
Due to the unstable nature of monolayer, it is very difficult to realize
the industrial applications of monolayer phosphorene.
However, successful efforts have
made it possible to fabricate experimentally high-quality monolayer
phosphorene using a controlled thinning process with transmission
electron microscopy and subsequent performance of atomic-resolution
imaging~\cite{Lee.NL.20:559}. Likewise, phosphorene
can also be synthesized experimentally using several techniques, including
liquid exfoliation and mechanical cleavage~\cite{Li-NN.9:372,Lu-NR.7:853}.
It has been shown that spin-orbit interaction~\cite{Kurpas-PRB.94:155423,Sattari-MSEB.278:115625,Luo-OE.28:9089,Farzaneh-PRB.100:245429,Popovic-PRB.92:035135}
and inversion symmetry breaking~\cite{Low-PRB.92:235447} crucially affect the electronic properties
of phosphorene.
Anisotropy in the band structure is a characteristic feature of
phosphorene, leading to its perspective optical, magnetic, mechanical
and electrical properties~\cite{Fei-NL.14:2884,Wang-NN.10:517,Hu-PRB.97:045209,Elahi-PRB.91:115412}.
Interesting transport properties as such electrical
conductivity~\cite{Sultana-JPCS.176:111257} and second
order nonlinear Hall effect~\cite{Yar-JPCM.35:165701}
in monolayer phosphorene have been investigated.
Novel applications of this quantum material have been envisioned in transistors,
batteries, solar cells, disease theranostics, actuators, thermoelectrics, gas sensing, humidity sensing,
photo-detection, bio-sensing, and ion-sensing devices~\cite{Tareen-PSSC.65:100336}.
Due to high carrier mobility and anisotropic in-plane properties, phosphorene is an appealing candidate for promising applications in nanoelectronics and nanophotonics~\cite{Ling-PNAS.112:4523,Churchill-NN.9:330,Liu-CSR.44:2732}.\\
On the other hand, the intriguing feature of quantum mechanics in
lattice systems is the Bloch oscillation of a particle in the
periodic potential of a perfect crystal lattice subjected
to a constant external force~\cite{Bloch.ZP.52:555,Zener.PRSL.145:523}.
It shows coherent dynamics of quantum
many-body systems~\cite{Ashcroft-Book}, originated
from the translational symmetry of crystals. It has been shown
that these oscillations appear with a
fundamental period that a semiclassical wave packet takes
to traverse a Brillouin-zone loop.
Analysis shows that Bloch oscillations in two superposed
optical lattices can split, reflect, and recombine matter
waves coherently~\cite{Pagel-PRA.102:053312}.
It was found that Wannier-Stark states(WS states) exhibit Bloch oscillations
with irregular character for irrational
directions of the static field in a tilted honeycomb
lattice within the tight-binding approximation~\cite{Kolovsky-PRA.87:3112}.
Theoretical study reveals that Berry curvature crucially
modifies the semiclassical dynamics of a system and affects the Bloch
oscillations of a wave packet under a constant external force, leading to a net drift of
the wave packet with time. Interestingly, loss of information
about the Berry curvature due to the complicated
Lissajous-like figures can be recovered via a time-reversal protocol.
For experimental measurement, a general technique for mapping the local
Berry curvature over the Brillouin zone
in ultracold gas experiments has been proposed~\cite{Kolovsky-PRA.87:3112}.
Bloch oscillations can be observed in semiconductor superlattices~\cite{Feldmann-PRB:46:7252}, ultracold atoms and Bose-Einstein condensates~\cite{Dahan-PRL.76:4508,Anderson-Sci:282:1686,Battesti-PRL.92:253001,Zhang-Op.4:571,Morsch-PRL.87:140402}, photonic structures~\cite{Pertsch-PRL.83:4752,Morandotti-PRL.83:4756,Sapienza-PRL.91:263902,Trompeter-PRL.96:023901,Trompeter-PRL.96:053903} and plasmonic waveguide arrays~\cite{Block-NC.5:3843}.
Moreover, Bloch oscillations with periodicity to be an integer
multiple of the fundamental period have been reported~\cite{Hoeller-PRB:98:024310}.
It is emphasized that Bloch oscillations essentially rely
on the periodicity of crystal quasimomentum, as well as
the existence of an energy gap, where both are the basic features of a
quantum theory of solids. From a semiclassical point of view,
Bloch oscillations are originated from the dynamics of a wave
packet formed from a single band. Using the acceleration
theorem~\cite{Nenciu-PLA:78:101}, the fundamental period
($T$) of this oscillation is determined to be the time taken by a wave packet in
traversing a loop across the Brillouin torus given by $T =\hbar|\textbf{G}|/\textbf{F}$,
with $\textbf{G}$ being the smallest reciprocal
vector parallel to a time-independent driving force $\textbf{F}$.
Fundamental Bloch oscillations may also be realized as a
coherent Bragg reflection originated from the discrete
translational symmetry of a lattice~\cite{Ashcroft-Book}.
Remarkably, Bloch oscillation based methods are effectively
used in cold-atom applications, such as for precision measurements
of the fine-structure constant~\cite{Clade-PRL:96:033001},
gravitational forces~\cite{Anderson-Sci:282:1686,Roati-PRL:92:230402},
even on very small length scales~\cite{Ferrari-PRL:97:060402}.
Bloch dynamics has been studied in many condensed matter systems,
for instance, lattices with long-range hopping~\cite{Stockhofe-PRA:91:023606},
two-dimensional lattices~\cite{Witthaut-NJP:6:41}, two-dimensional optical lattices~\cite{Kolovsky-PRA.67:063601,Price-PRA.85:033620},
Weyl semimetals~\cite{Wang-PRA.94:031603}, beat note
superlattices~\cite{Masi-PRL.127:020601}, etc.
Recently, the experimental simulation of anyonic Bloch
oscillations using electric circuits has been reported~\cite{Zhang-NC.13:2392}.\\
In this paper, we investigate Bloch dynamics in monolayer phosphorene with broken
inversion symmetry. We find that the wave packet exhibits Bloch oscillations
that strongly depend on the band structure of the system.
It is shown that spin-orbit interaction has remarkable
effect on the Bloch dynamics. The dynamics
is modified considerably under the influence of an in-plane
electric and transverse magnetic fields.\\
The paper is organized as follows: In Sec.~\ref{Sec:Methodology}, the
tight-binding Hamiltonian of a monolayer phosphorene with
broken inversion symmetry is presented. The Hamiltonian is reduced
to a two band system at the high symmetry point $\Gamma$,
followed by the determination
of eigenstates, eigenvalues and the Berry curvature.
The dynamical equations are presented in this section.
\newline
Sec.~\ref{Sec:RandD} contains the investigation of Bloch oscillations
in monolayer phosphorene with broken inversion symmetry.
The effects of spin-orbit interaction on the Bloch dynamics are presented.
Morever, the effects of in-plane electric and transverse magnetic
fields are demonstrated in this section.  \newline
Finally, conclusions are drawn in Sec.~\ref{Sec:Conc}.\\
\section{Methodology}\label{Sec:Methodology}
\label{Sec:Methodology}
In this section, we present the model and related
theoretical background of the work.
\subsection{Theory and Model}
\begin{figure}[!ht]
\centerline{\psfig{figure=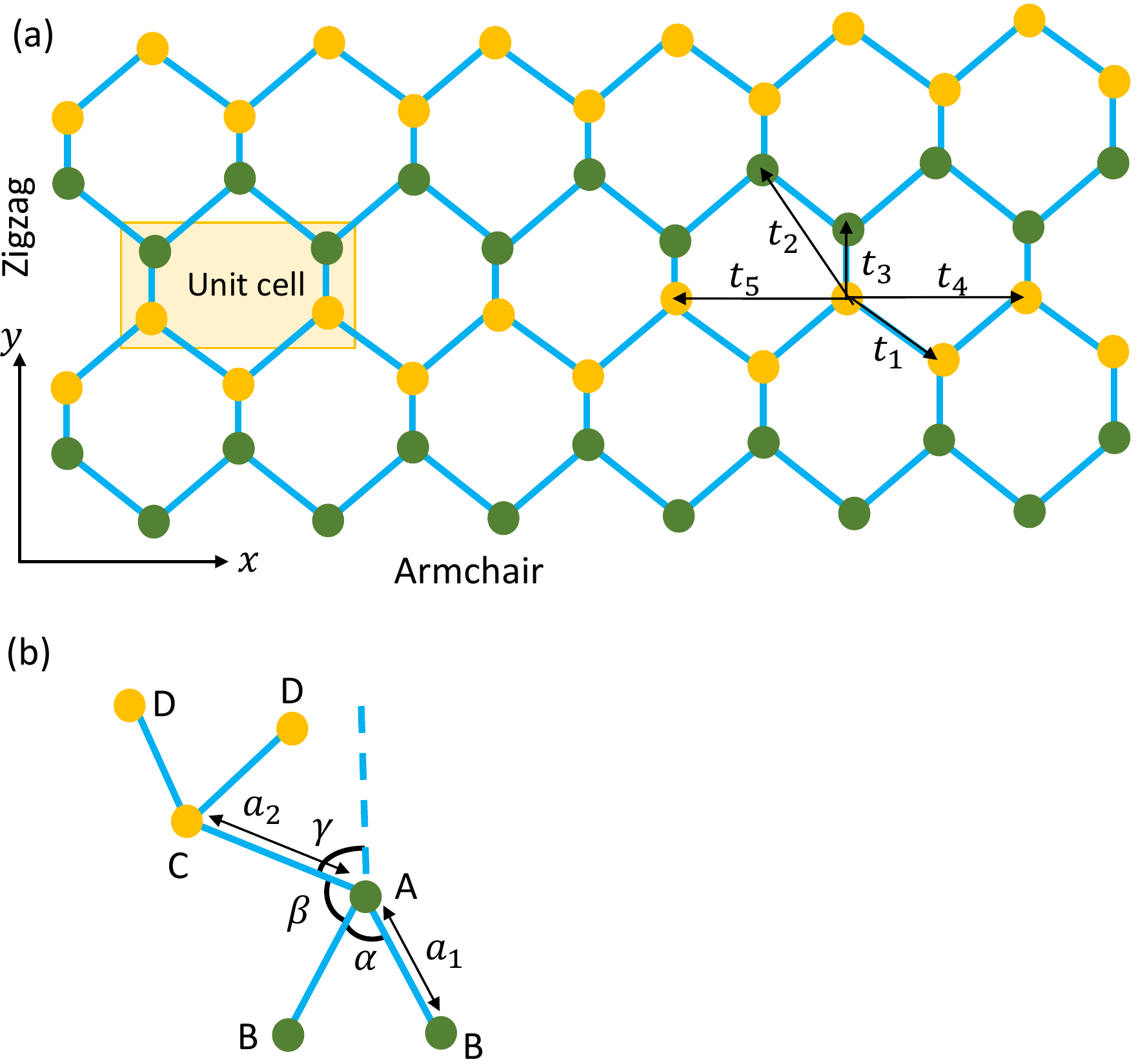,width=\columnwidth}}
\caption{Schematic realization of the lattice structure of monolayer
phosphorene in which the shaded circles with different colors represent atoms positioned in
different planes within a single puckered layer.
The shaded region shows the unit cell containing four atoms, wheras
the factors $t_1,t_2,t_3,t_4,t_5$ characterize the five hopping parameters
among the nearest neighbors in the tight binding model.
(b) Schematic visualization of the bond lengths and bond angles in monolayer phosphorene.}
\label{Figure1}
\end{figure}

We consider the band structure of black phosphorus (phosphorene)
with a spin-independent tight-binding model using a basis of $s$
orbital and three $p$ orbitals. The unit cell of phosphorene
consists of four phosphorus atoms, see Fig.~\ref{Figure1} (a), leading
to the formation of sixteen bands. The band structure with band gap
of monolayer phosphorene can be determined by evaluating the
hopping energy and overlaps between neighboring atoms, indexing
the symmetries of eigenstates at the $\Gamma$ point. In general,
the wave functions constructed in this way consist of $sp^3$
hybridized atomic orbitals.
Using the method of tight-binding model,
the Hamiltonian for monolayer phosphorene with broken inversion symmetry can be described
as~\cite{Pereira-PRB.92:075437,Rudenko-PRB.89:201408}
\begin{align}\label{eq:Hamiltonian4by4}
\mathcal{H}_0\left(\textbf{k}\right) =
\begin{pmatrix}
  u_A+\Delta & t_{AB}\left(k\right) & t_{AD}\left(k\right) & t_{AC}\left(k\right)\\
  t_{AB}\left(k\right)^\ast & u_B+\Delta  & t_{AC}\left(k\right)^\ast & t_{AD}\left(k\right)\\
  t_{AD}\left(k\right)^\ast & t_{AC}\left(k\right) & u_D-\Delta  & t_{AB}\left(k\right)\\
  t_{AC}\left(k\right)^\ast & t_{AD}\left(k\right)^\ast & t_{AB}\left(k\right)^\ast & u_C-\Delta
\end{pmatrix},
\end{align}
with eigenvectors $\left[\psi_A \ \psi_B\ \psi_D\ \psi_C\right]^T$
and $u_A$, $u_B$, $u_C$, and $u_D$ are the on-site energies, which are
taken as $U$, with the $A-D$ subscripts characterizing the four sublattice
labels shown in Fig.~\ref{Figure1}.
Moreover, $t_{AB}\left(k\right)$, $t_{AC}\left(k\right)$,
and $t_{AD}\left(k\right)$ denote the coupling factors.
Considering the $C_{2h}$ group symmetry of the black
phosphorus lattice structure~\cite{Ezawa-NJP.16:115004}
and $t_{AD}\left(k\right)^\ast=t_{AD}\left(k\right)$,
a reduced two-band Hamiltonian for monolayer phosphorene in the
vicinity of the Fermi level can be obtained as~\cite{Pereira-PRB.92:075437}
\begin{align}\label{eq:Hamiltonian2by2}
\mathcal{H}_0\left(\textbf{k}\right) =
\begin{pmatrix}
  U+ t_{AD}\left(k\right)+\Delta & t_{AB}\left(k\right)+t_{AC}\left(k\right)\\
  t_{AB}\left(k\right)^\ast+t_{AC}\left(k\right)^\ast & U+ t_{AD}\left(k\right)-\Delta
\end{pmatrix},
\end{align}
where
\begin{align}\label{eq:tLM}
t_{AB}\left(k\right)&=2t_1\cos\left[k_x a_1\sin\left(\alpha/2\right)\right]
e^{-ik_ya_1\cos\left(\alpha/2\right)}\notag\\&
+2t_3\cos\left[k_x a_1\sin\left(\alpha/2\right)\right]
e^{ik_y\left[a_1\cos\left(\alpha/2\right)+2a_2\cos\gamma\right]},
\end{align}
\begin{align}\label{eq:tLN}
t_{AC}\left(k\right)&=t_2e^{ik_ya_2\cos\beta}
+t_5e^{-ik_y\left[2a_1\cos\left(\alpha/2\right)+a_2\cos\gamma\right]},
\end{align}
\begin{align}\label{eq:tLQ}
t_{AD}\left(k\right)&=4t_4\cos\left\{k_y\left[ a_1\cos\left(\alpha/2\right)+a_2\cos\gamma\right]\right\}\notag\\&\times
\cos\left[k_x a_1\sin\left(\alpha/2\right)\right],
\end{align}
where the bond length, $a_1=2.22\ \textrm{\AA}$ represents
the distance between nearest-neighbor sites in
sublattices $A$ and $B$ or $C$ and $D$ and $a_2=2.24\ \textrm{\AA}$
is the distance between nearest-neighbor sites in sublattices $A$
and $C$ or $B$ and $D$; the bond angles are $\alpha=96^{\circ},5^{\circ}$,
$\beta=101^{\circ},9^{\circ}$, $\cos\gamma=-\cos\beta/\cos\alpha$
as shown in Fig.~\ref{Figure1} (b),
whereas $t_1=-1.220\ \textrm{eV}$, $t_2=3.665\ \textrm{eV}$,
$t_3=-0.205\textrm{eV}$, $t_4=-0.105\ \textrm{eV}$, and
$t_5=-0.055\ \textrm{eV}$, see Fig.~\ref{Figure1} (a), are
the corresponding hopping parameters for nearest-neighbor
couplings~\cite{Rudenko-PRB.89:201408}. Using Eq.~\eqref{eq:Hamiltonian4by4}, solution of the secular
equation leads to the energy dispersion in the form
\begin{align}\label{eq:spectrum}
&E_\lambda\left(k_x,k_y\right)=U+t_{AD}\notag\\&
+\lambda\sqrt{\left(t_{AB}+t_{AC}\right)\left(t_{AB}+t_{AC}\right)^\ast+\Delta^2},
\end{align}%
where $\lambda=\pm 1$ is the band index, with the positive sign showing
the conduction band and negative sign characterizes the valence band.
Hence, expanding the structure factors in the vicinity of $\Gamma$ point
and retaining the terms up to second order in $k$, the
two-band Hamiltonian of monolayer phosphorene with broken inversion
symmetry within the long-wavelength approximation
can be obtained as~\cite{Yar-JPCM.35:165701}
\begin{align}\label{eq:Hamiltonian}
\mathcal{H}_0\left(\textbf{k}\right) &= \left(u_0+\eta_x k^2_x
+\eta_y k^2_y\right)\mathbb{1}+\left(\delta+\gamma_x k^2_x
+\gamma_y k^2_y\right)\sigma_x\notag\\&
-\chi k_y\sigma_y+\sigma_z\Delta,
\end{align}%
where $u_0=-0.42\ \textrm{eV}$, $\delta= 0.76\ \textrm{eV}$,
$\eta_x=0.58\ \textrm{eV}\textrm{\AA}^2$, $\eta_y = 1.01\ \textrm{eV}\textrm{\AA}^2$,
$\gamma_x = 3.93\ \textrm{eV}\textrm{\AA}^2$, $\gamma_y = 3.83\ \textrm{eV}\textrm{\AA}^2$,
and $\chi = 5.25\ \textrm{eV}\textrm{\AA}$ are the band parameters
which remain the same as used in Ref.~\cite{Pereira-PRB.92:075437} and they include the
contribution from the five-hopping energies of the tight-binding
model for a BP sheet and its lattice geometry as shown in Fig.~\ref{Figure1}.
In Eq.~\eqref{eq:Hamiltonian}, $k_x$ and $k_y$ are the in-plane crystal momenta, whereas
$\sigma_x$, $\sigma_y$, and $\sigma_z$ represent the $2\times 2$ Pauli matrices
and $\mathbb{1}$ stands for the unit matrix.
Moreover, $\Delta$ denotes the broken inversion symmetry
induced band gap in the energy spectrum of the system.
The energy dispersion of monolayer phosphorene is
\begin{align}\label{eq:spectrum}
E_\lambda\left(k_x,k_y\right) &=\in_1+\lambda\sqrt{\in^2_2+\Delta^2},
\end{align}%
where we have defined: $\in_1\equiv u_0+\eta_x k^2_x+\eta_y k^2_y$,
$\in_3\equiv \delta+\gamma_x k^2_x+\gamma_y k^2_y$,
$\in_4\equiv \chi k_y$,
$\in_2\equiv \sqrt{\in^2_3+\in^2_4}$.
The first term in the right hand side of Eq.~\eqref{eq:spectrum}
makes the band structure of phosphorene highly anisotropic.
The Hamiltonian in Eq.~\eqref{eq:Hamiltonian} can be diagonalized
using the standard diagonalization method.
Consequently, using the polar notation, normalized eigenstates of the
aforementioned Hamiltonian are described as
\begin{equation}\label{eq:States}
\psi_\lambda\left(k_x,k_y\right)=
\frac{e^{i\textbf{k}\cdot\textbf{r}}}{\sqrt{2S}}\left(
    \begin{array}{ccc}
     \sqrt{1+\lambda\cos\theta_k} \\ \\
      e^{-i\varphi_k}\lambda\sqrt{1-\lambda\cos\theta_k}
    \end{array}
  \right),
\end{equation}%
with $S$ being the dimensions of the system, $\tan\varphi_k=\frac{\in_4}{\in_3}$, and $\tan\theta_k=\frac{\in_2}{\Delta}$.\\
The inversion symmetry breaking in monolayer phosphorene leads to a finite Berry
curvature. Such curvature in momentum space can be evaluated
using Eqs.~\eqref{eq:spectrum} and~\eqref{eq:States} in the vicinity of the $\Gamma$ point as~\cite{Yar-JPCM.35:165701}
\begin{align}\label{eq:BerryCurvature}
&\Omega_\lambda\left(\textbf{k}\right)=\lambda\chi\gamma_x k_x\Delta\notag\\&\times\frac{\left[\left(\in_3-\frac{\gamma_y\in^2_4}{\chi^2}\right)
\left(\in_3-\frac{3\gamma_y\in^2_4}{\chi^2}\right)-\in^2_4\left(1
+\frac{3\gamma^2_y\in^2_4}{\chi^4}\right)\right]}{\left(\in^2_3+\in^2_4
+\Delta^2\right)^{3/2}\left(\in^2_3+\in^2_4\right)}.
\end{align}
It is illustrated that the Berry curvatures of the conduction
($\lambda=+$) and valence ($\lambda=-$) bands have opposite
signs and vanish in the absence of
the band gap induced in the energy spectrum. The Berry
curvature exhibits very interesting symmetry properties~\cite{Yar-JPCM.35:165701}.
\subsection{Semiclassical Dynamics of Wave Packet}
\label{Sec:SDWP}
We develop formalism for semiclassical dynamics
of a particle in monolayer phosphorene with broken inversion symmetry.
We consider a single particle that is prepared in a wave
packet state having center of mass at position $\textbf{r}$ with momentum
$\textbf{k}$~\cite{Ashcroft-Book,Dahan-PRL.76:4508}.
The Bloch velocity of a wave packet can be described as
\begin{align}\label{eq:SCv}
\dot{\textbf{r}}_\lambda=\frac{1}{\hbar}\pmb{\nabla}_k E_\lambda
\left(\textbf{k}\right)-\left(\dot{\textbf{k}}\times\textbf{e}_z\right)\Omega_\lambda\left(\textbf{k}\right),
\end{align}%
with
\begin{align}\label{eq:SCF}
\hbar\dot{\textbf{k}}=\textbf{F},
\end{align}%
where $\textbf{e}_z$ is the unit vector in the $z$-direction, the
first term on the right hand side of Eq.~\eqref{eq:SCv} denotes the group velocity
evaluated by taking the gradient of energy spectrum in momentum space
and the second term describes the Berry velocity.
Eq.~\eqref{eq:SCv} shows that the electron band
velocity is periodic in crystal momentum $k$.
It has been found that the effects of Berry curvature
can also be determined in the semiclassical dynamics of a wave packet in
a time-dependent one-dimensional (1D) optical lattice~\cite{Xiao-RMP.82:1959,Kitagawa-PRB.82:235114,Pettini-PRA.83:013619}
which is defined over a 2D parameter space, composed of the one-dimensional
quasimomentum and time. The Bloch oscillations of a wave packet in such a potential
have been investigated in Ref.~\cite{Pettini-PRA.83:013619}.\\
We evaluate the Bloch velocity of the wave packet
in the conduction band using Eqs.~\eqref{eq:spectrum},~\eqref{eq:BerryCurvature},  and~\eqref{eq:SCv}.
As a consequence, the $x$-component of the velocity acquires the form
\begin{align}\label{eq:vx}
v_x\left(\bf{k}\right)&=-\frac{4d_1t_4}{\hbar}g_1\left(k\right)-\frac{2d_1}{\hbar}
\left\{4g_2\left(k\right)+g_3\left(k\right)+
4g_4\left(k\right)\right.\notag\\& \left.
+4g_5\left(k\right)+\Delta^2\right\}^{-1/2}
\left\{g_6\left(k\right)+ g_7\left(k\right)
+g_8\left(k\right)\right\}\notag\\&
+\frac{F_y}{\hbar}\Omega_\lambda\left(\textbf{k}\right),
\end{align}%
and the $y$-component is
\begin{align}\label{eq:vy}
v_y\left(\bf{k}\right)&=-\frac{4d_1t_4}{\hbar}g_9\left(k\right)-\frac{2d_1}{\hbar}
\left\{4g_2\left(k\right)+g_3\left(k\right)+
4g_4\left(k\right)\right.\notag\\& \left.
+4g_5\left(k\right)+\Delta^2\right\}^{-1/2}
\left\{4g_{10}\left(k\right)+ g_{11}\left(k\right)
+g_{12}\left(k\right)\right.\notag\\& \left.
+g_{13}\left(k\right)\right\}
-\frac{F_x}{\hbar}\Omega_\lambda\left(\textbf{k}\right),
\end{align}%
where we have defined:
\begin{align}\label{eq:funcs}
&g_1\left(k\right)=\sin\left(k_xd_1\right)\cos\left(k_yd_2\right),\notag\\& g_2\left(k\right)=\left[t^2_1+t^2_3+2t_1t_3\cos\left(2k_yd_2\right)\right]\cos^2\left(k_xd_1\right),\notag\\&
g_3\left(k\right)=t^2_2+t^2_5+2t_2t_5\cos\left(2k_yd_2\right),\notag\\&
g_4\left(k\right)=t_3\left[t_2\cos\left(k_yd_2\right)+t_5\cos\left(3k_yd_2\right)\right]\cos\left(k_xd_1\right),\notag\\&
g_5\left(k\right)=t_1\left(t_2+t_5\right)\cos\left(k_xd_1\right)\cos\left(k_yd_2\right),\notag\\&
g_6\left(k\right)=\left[t^2_1+t^2_3+2t_1t_3\cos\left(2k_yd_2\right)\right]\sin\left(2k_xd_1\right),\notag\\&
g_7\left(k\right)=t_3\left[t_2\cos\left(k_yd_2\right)+t_5\cos\left(3k_yd_2\right)\right]\sin\left(k_xd_1\right),\notag\\& g_8\left(k\right)=t_1\left(t_2+t_5\right)\sin\left(k_xd_1\right)\cos\left(k_yd_2\right),\notag\\&
g_9\left(k\right)=\cos\left(k_xd_1\right)\sin\left(k_yd_2\right),\notag\\&
g_{10}\left(k\right)=t_1t_3\sin\left(2k_yd_2\right)\cos^2\left(k_xd_1\right),\notag\\&
g_{11}\left(k\right)=t_2t_5\sin\left(2k_yd_2\right),\notag\\&
g_{12}\left(k\right)=t_3\left[t_2\sin\left(k_yd_2\right)+t_3t_5\sin\left(3k_yd_2\right)\right]\cos\left(k_xd_1\right),\notag\\&
g_{13}\left(k\right)=t_1\left(t_2+t_5\right)\cos\left(k_xd_1\right)\sin\left(k_yd_2\right).
\end{align}%
Eqs.~\eqref{eq:vx} and~\eqref{eq:vy} reveal that the Bloch velocities
$v_x\left(\textbf{k}\right)$ and $v_y\left(\textbf{k}\right)$ exhibit oscillatory
behaviour over the entire range of $k_x$ and $k_y$. It is illustrated
that both $v_x\left(\textbf{k}\right)$ and $v_y\left(\textbf{k}\right)$
consist of group and Berry velocities which can be separated as
\begin{align}\label{eq:GroupVelocity}
\textrm{v}_\textbf{k}\left(+\textbf{F}\right)+\textrm{v}_\textbf{k}\left(-\textbf{F}\right)=
\frac{2}{\hbar}\frac{\partial E\left(\textbf{k}\right)}{\partial\textbf{k}}.
\end{align}%
\begin{align}\label{eq:BerryVelocity}
\textrm{v}_\textbf{k}\left(+\textbf{F}\right)-\textrm{v}_\textbf{k}\left(-\textbf{F}\right)=
-\frac{2}{\hbar}\left(\textbf{F}\times\textbf{e}_z\right)\Omega\left(\textbf{k}\right).
\end{align}%
This transformation is equivalent to a time-reversal operation,
and it obviously removes the effects of the complex Lissajous-like
figures in 2D. Interesting behaviours are exhibited by the Bloch
velocity $v\left(k_x,ky\right)$ in the Brillouin zone. In particular,
the $x$-component of the group velocity, $v_{x}\left(k_x,ky\right)$,
vanishes at $k_x=0,\ k_y\neq 0$
as is clear from Eq.~\eqref{eq:GroupVelocity},
whereas the $y$-component, $v_{y}\left(k_x,ky\right)$, remains finite.
Likewise, $v_{y}\left(k_x,ky\right)$
vanishes at $k_y=0,\ k_x\neq 0$, $v_{x}\left(k_x,ky\right)$ remains finite.
Further, $v_{x}\left(k_x,ky\right)$ changes its sign by changing the sign
of $k_x$, whereas $v_{y}\left(k_x,ky\right)$ changes its sign with $k_y$.
Moreover, the group velocity is affected by the band gap opened
in the energy spectrum due to the broken inversion symmetry,
however, it remains finite even if the aforementioned
symmetry is retained. In contrast, Berry velocity depends on
the inversion symmetry breaking which becomes zero if the
system preserves the inversion symmetry. The Berry velocity in Eq.~\eqref{eq:BerryVelocity},
$v_{j}\left(k_x,ky\right)$ with $j=x,y$,
exhibits the following symmetry properties:
(i) The Berry velocity shows mirror reflection symmetry $k_y\leftrightarrow-k_y$, i.e., $v_{j}\left(k_x,-k_y\right)=v_{j}\left(k_x,k_y\right)$.
(ii) It remains finite in a crystal system with broken inversion
symmetry, i.e., a crystal lattice with inversion symmetry requires $v_{j}\left(\textbf{k}\right)=v_{j}\left(-\textbf{k}\right)=0$.
(iii) It shows the character of an odd function in
momentum space, i.e., $v_{j}\left(-k_x,k_y\right)=-v_{j}\left(k_x,k_y\right)$,
reflecting time-reversal symmetry of the system. (iv) It changes sign
when the direction of the applied force is reversed.\\
\section{Results and Discussion on Bloch Dynamics}\label{Sec:RandD}
\begin{figure}[!ht]
\centerline{\psfig{figure=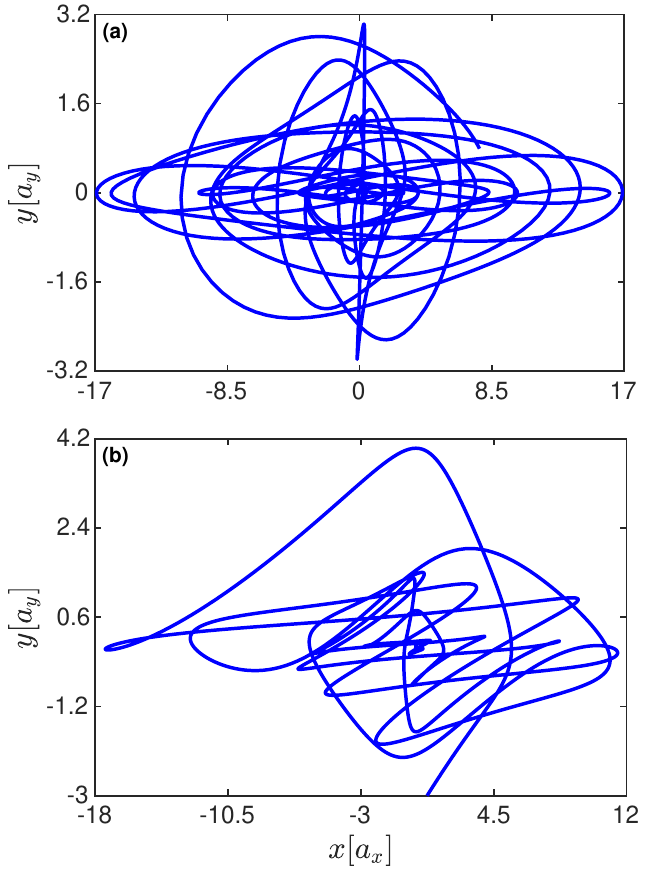,width=\columnwidth}}
\caption{Lissajous-like figure for a wave packet in inversion symmetry
broken monolayer phosphorene using the ratio: (a)
$F_x : F_y= 2F_0 : 12F_0$ and (b) $F_x : F_y= 12F_0 : 2F_0$ with $F_0 = \delta/a_x$.
Parameters used in the numerical simulations are: $a_x \equiv a_1= 3.32\textrm{\AA}$;
$a_y \equiv a_2= 4.38\textrm{\AA}$; $\Delta=\delta$ and other parameters
are the same as given in the text.}
\label{Figure2}
\end{figure}
In this section, we present the results on Bloch oscillations in
monolayer phosphorene with broken inversion symmetry.
For analyzing the remarkable feature of dimensionality, we plot the
real-space trajectories of the Bloch oscillations in Fig.~\ref{Figure2} which
reveals Lissajous-like oscillations. It has been shown that
1D Bloch oscillations in the presence of separable
potentials are simply superposed along the $x$ and $y$-axes. The wave packet dynamics
exhibits periodic behaviour along $k_i$ with periods $T_j= h/|F_j|a$ for an arbitrary
force $F = \left(F_x,F_y \right)$. The resulting dynamics
depends on the ratio $F_x : F_y$. For nonseparable potentials,
similar dynamical behavior can be expected when the applied force is weak and
Landau-Zener tunneling is negligibly small~\cite{Kolovsky-PRA.67:063601,Witthaut-NJP.6:41}.
\begin{figure}[!ht]
\centerline{\psfig{figure=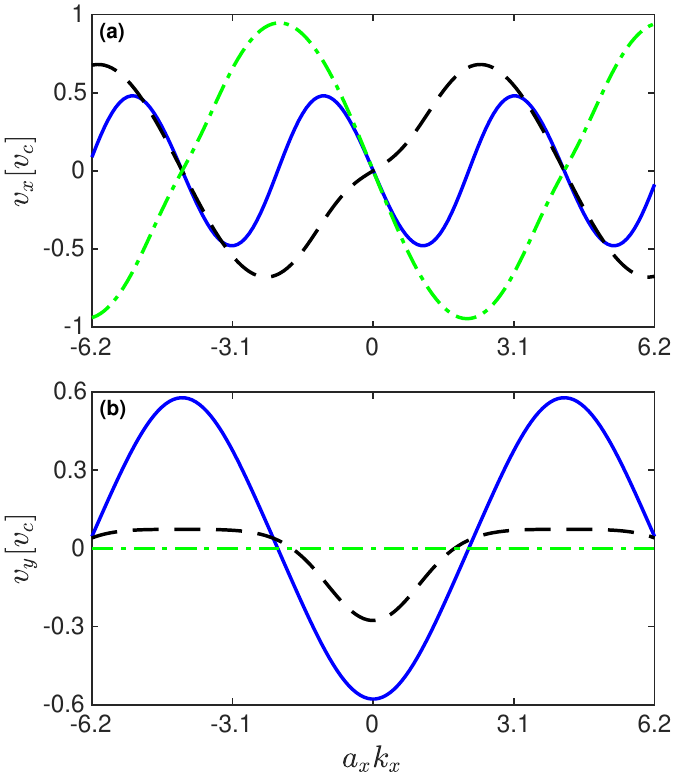,width=\columnwidth}}
\caption{Group velocity of charge carrier in inversion symmetry
broken monolayer phosphorene in units of $v_c$, with $v_c=\chi/\hbar$
being the characteristic velocity, versus the crystal momentum $k_x$.
Panel (a) shows the group velocity in the $x$-direction, whereas (b) in the $y$-direction.
In each panel, the blue solid curve is used for $k_yd_2=\frac{\pi}{2}$, black dashed curve for $k_yd_2=\frac{\pi}{4}$, and green dash-dotted curve for $k_yd_2=\pi$.
The parameters used in the numerical simulations are the same as used for Fig.~\ref{Figure2}.}
\label{Figure3}
\end{figure}
The real-space Lissajous-like figures describe complicated
two-dimensional oscillations, which are bounded by $x_{j} \propto v_{j}T_{j}$,
see Fig.~\ref{Figure2}. Note that we have adopted a scheme in
which the ratio $F_x : F_y$ has been made large, where the
Bloch electron covers a large area of the Brillouin zone
during a single Bloch oscillation. It is obvious that the
Lissajous-like figure is approximately bounded by the Bloch
oscillation lengths, and so it makes the effects of
Berry curvature ambiguous within the bounded region.\\
This trajectory can be changed significantly by the Berry curvature,
if we wait until the wave packet drifts outside the bounded region.
As a consequence, only the net Berry curvature encountered along a path will be
measured in experiments. Information regarding the distribution
of Berry curvature in momentum space will be lost, in particular,
whether its sign changes.
Moreover, an additional drift in the
position of wave packet may occur in 2D, independent of the Berry curvature, if
the wave packet does not start at high symmetry points such as
the zone center $k_0 = (0,0)$~\cite{Mossmann-JPAMG.38:3381,Zhang-PRA.82:025602}.
Hence, merely the observation of a transverse drift in the position
of wave packet is not a conclusive evidence of a finite Berry curvature.\\
To better understand the Bloch dynamics in monolayer
phosphorene, the group velocity of the Bloch electron as a function of
crystal momentum $k_x$ is plotted in Fig.~\ref{Figure3}. This figure shows
that the group velocity of the Bloch electron is well pronounced in the
Brillouin zone that strongly depends on the initial momentum $k_y$ as is
obvious from comparison of the blue solid, black dashed, and green dash-dotted
curves. In particular, the change in initial crystal momentum $k_y$ leads to the
change of phase and amplitude of oscillations. Comparison of panels (a) and (b) shows that the
group velocities $v_x$ and $v_y$ exhibit different dynamical behaviour, where the latter
vanishes at $k_y=\pi/d_2$. Further, the oscillation frequency and amplitude
of oscillations of the two components are also very different.
\begin{figure}[!ht]
\centerline{\psfig{figure=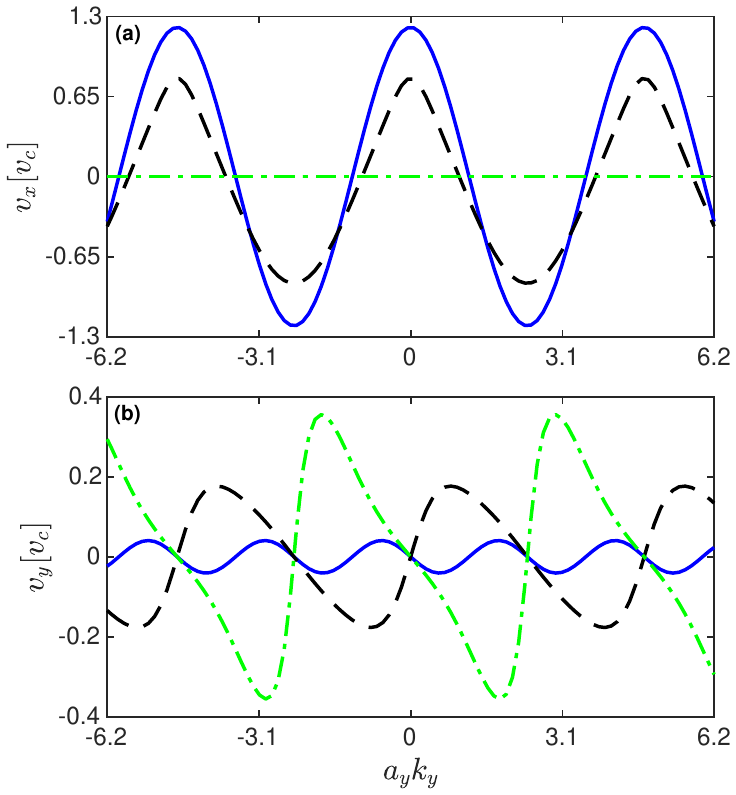,width=\columnwidth}}
\caption{Group velocity of charge carrier in inversion symmetry
broken monolayer phosphorene versus the crystal momentum $k_y$.
Panel (a) shows the group velocity in the $x$-direction, whereas (b) in the $y$-direction.
In each panel, the blue solid curve is used for $k_xd_1=\frac{\pi}{2}$, black dashed curve for $k_xd_1=\frac{\pi}{4}$, and green dash-dotted curve for $k_xd_1=\pi$
and other parameters remain the same as used for Fig.~\ref{Figure2}.}
\label{Figure4}
\end{figure}
For more insight, we show the group velocity of the Bloch electron
as a function of the crystal momentum $k_y$ in Fig.~\ref{Figure4}
for different values of the initial momentum $k_x$.
It mimics the behaviour of group velocity as shown in Fig.~\ref{Figure3}.
However in this case, the group velocity $v_x$ vanishes at $k_x=\pi/d_1$, see
panel (a), where $v_y$ remains finite, see panel (b).
Likewise, comparison of Figs.~\ref{Figure3} and~\ref{Figure4} reveals that the
oscillation frequency and amplitude of oscillations of the group
velocities are different as a function of $k_x$ and $k_y$, in particular, the
oscillation frequency of $v_y$ is large when it is analyzed as
a function of the crystal momentum $k_y$, see Figs.~\ref{Figure3} (b) and~\ref{Figure4} (b).
\begin{figure}[!ht]
\centerline{\psfig{figure=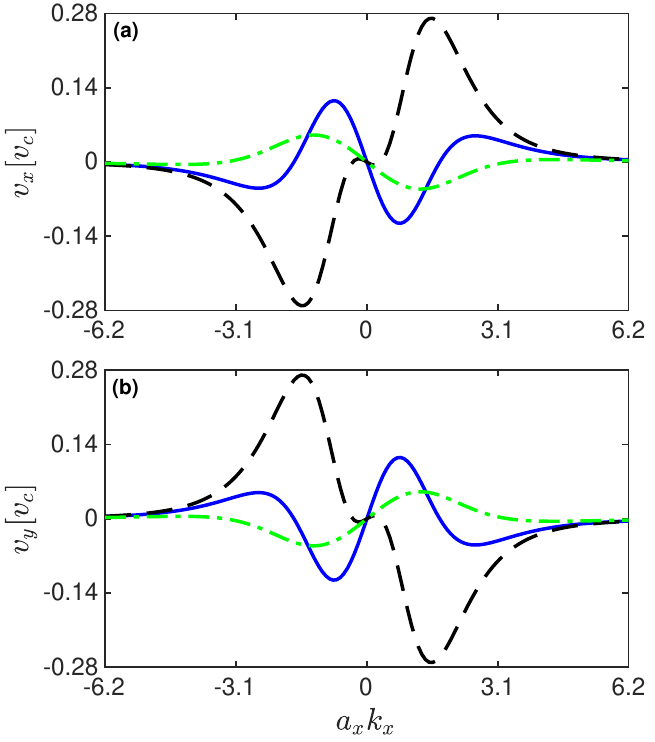,width=\columnwidth}}
\caption{Berry velocity of charge carrier in inversion symmetry
broken monolayer phosphorene versus the crystal momentum $k_x$.
Panel (a) shows the Berry velocity in the $x$-direction, whereas (b) in the $y$-direction.
In each panel, the blue solid curve is used for $k_yd_2=\frac{\pi}{2}$, black dashed curve for $k_yd_2=\frac{\pi}{4}$, and green dash-dotted curve for $k_yd_2=\pi$.
Other parameters used in the numerical
simulations are the same as used for Fig.~\ref{Figure2}.}
\label{Figure5}
\end{figure}
Moreover, we show the Berry velocity as a function of
crystal momentum $k_x$ in Fig.~\ref{Figure5} for different
values of the initial crystal momentum $k_y$.
Analysis of this figure
shows that the Berry velocity reflects the aforementioned
symmetry properties. In particular, comparison of
the blue solid, black dashed, and green dash-dotted curves in
both panels (a) and (b) shows that the $x$- and $y$-components
of the Berry velocity changes significantly
by changing the initial crystal momentum $k_y$, where the
change in amplitude and phase of oscillations can be seen.
Further, comparison of panels (a) and (b) shows that
the $x$- and $y$-components
of the Berry velocity oscillate with phase difference of $\pi$.
Interestingly, both components of the Berry velocity vanish at $k_x=0$
which are also negligibly small in the regions, $a_xk_x< -5$ and $a_xk_x> 5$
and well pronounced in the region, $-5<a_xk_x> 5$.
\begin{figure}[!ht]
\centerline{\psfig{figure=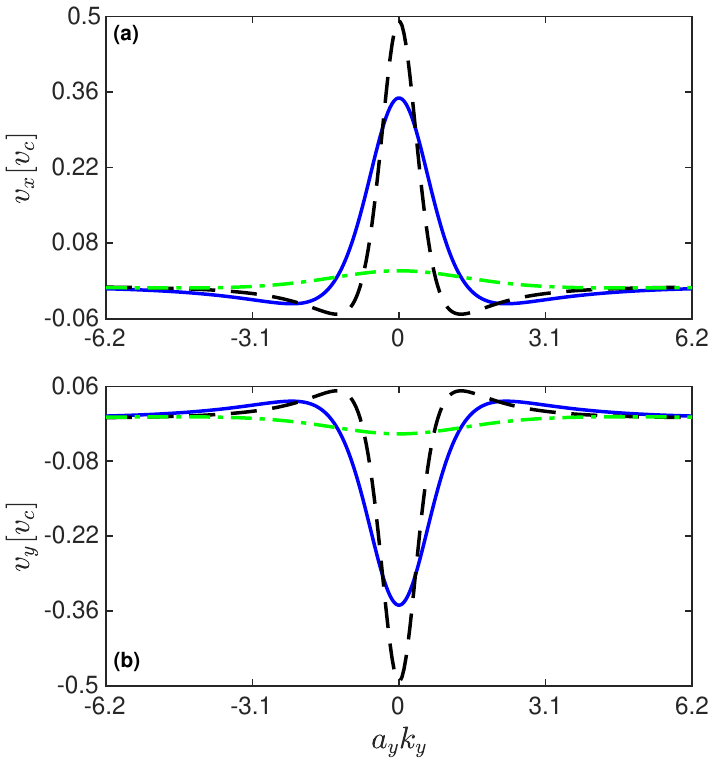,width=\columnwidth}}
\caption{Berry velocity of charge carrier in inversion symmetry
broken monolayer phosphorene versus the crystal momentum $k_y$.
Panel (a) shows the Berry velocity in the $x$-direction,
whereas (b) in the $y$-direction.
In each panel, the blue solid curve is used for $k_xd_1=\frac{\pi}{2}$,
black dashed curve for $k_xd_1=\frac{\pi}{4}$, and green
dash-dotted curve for $k_xd_1=\pi$.
Other parameters used in the numerical
simulations are the same as used for Fig.~\ref{Figure2}.}
\label{Figure6}
\end{figure}
For further understanding, the Berry velocity as a function of
crystal momentum $k_y$ is shown in Fig.~\ref{Figure6} for different
values of the initial crystal momentum $k_x$. In this case, the
Berry velocity exhibits interesting dynamical behaviour.
In particular, a single peak around $k_y\approx 0$ appears in
contrast to the former case when the Berry velocity is plotted as a
function of $k_x$ where two peaks are obtained on the left
and right of $k_x= 0$ with opposite phases. Moreover, the Berry
velocity vanishes in the regions, $a_yk_y\ll 0$ and $a_yk_y\gg 0$.
\subsection{Bloch dynamics in an in-plane electric field along $x$-axis}
\label{Sec:Hamiltonian}
In this case, the electric field is applied
in the $x$-direction, i.e., $E=E_x$, hence $E_y=0$. As a consequence,
$k_y(t)=k_y=\textrm{constant}$ and $k_x(t)=k_x(0)+\frac{eE_x}{\hbar} t$
that sweeps the entire Brillouin zone. After reaching the right end point $k_x=\pi/a_x$
electron is Bragg-reflected and continues from the left end point $k_x=-\pi/a_x$.
\begin{figure}[!ht]
\centerline{\psfig{figure=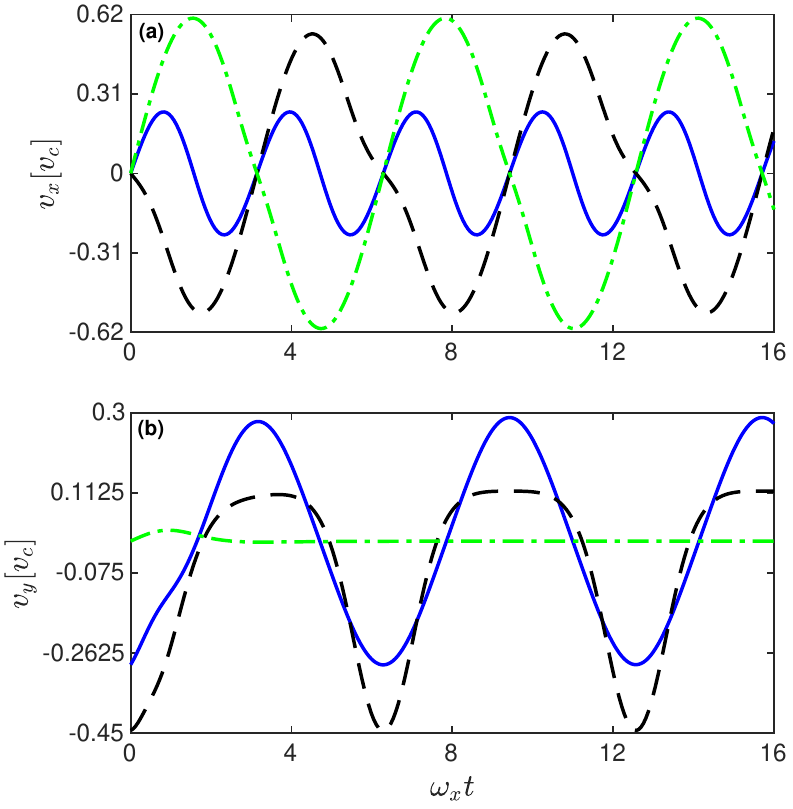,width=\columnwidth}}
\caption{Bloch velocity as a function of time for monolayer phosphorene with $\omega_x=\frac{eE_xd_1}{\hbar}$.
Panel (a) shows the Berry velocity in the $x$-direction, whereas (b) in the $y$-direction.
In each panel, the blue solid curve is used for $k_yd_2=\frac{\pi}{2}$, black dashed curve for $k_yd_2=\frac{\pi}{4}$, and green dash-dotted curve for $k_yd_2=\pi$.
Other parameters used in the numerical
simulations are the same as used for Fig.~\ref{Figure2}.}
\label{Figure7}
\end{figure}
\begin{figure}[!ht]
\centerline{\psfig{figure=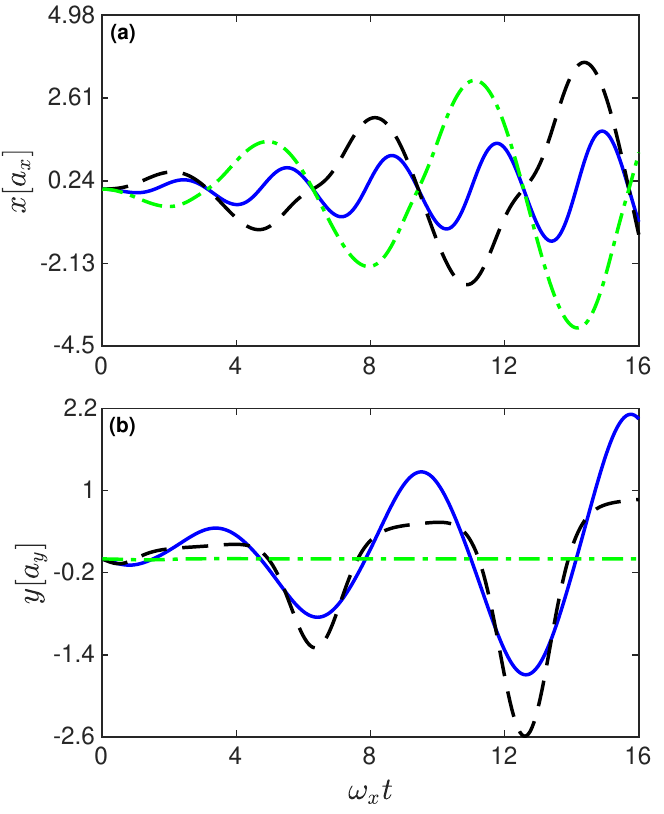,width=\columnwidth}}
\caption{Real space trajectories of Bloch particle as a function of time for monolayer phosphorene for: (a) $x$-component, (b) $y$-component, where in each panel, the blue solid curve is used for $k_yd_2=\frac{\pi}{2}$, black dashed curve for $k_yd_2=\frac{\pi}{4}$, and green dash-dotted curve for $k_yd_2=\pi$.
Other parameters used in the numerical
simulations are the same as used for Fig.~\ref{Figure2}.}
\label{Figure8}
\end{figure}
\begin{figure}[!ht]
\centerline{\psfig{figure=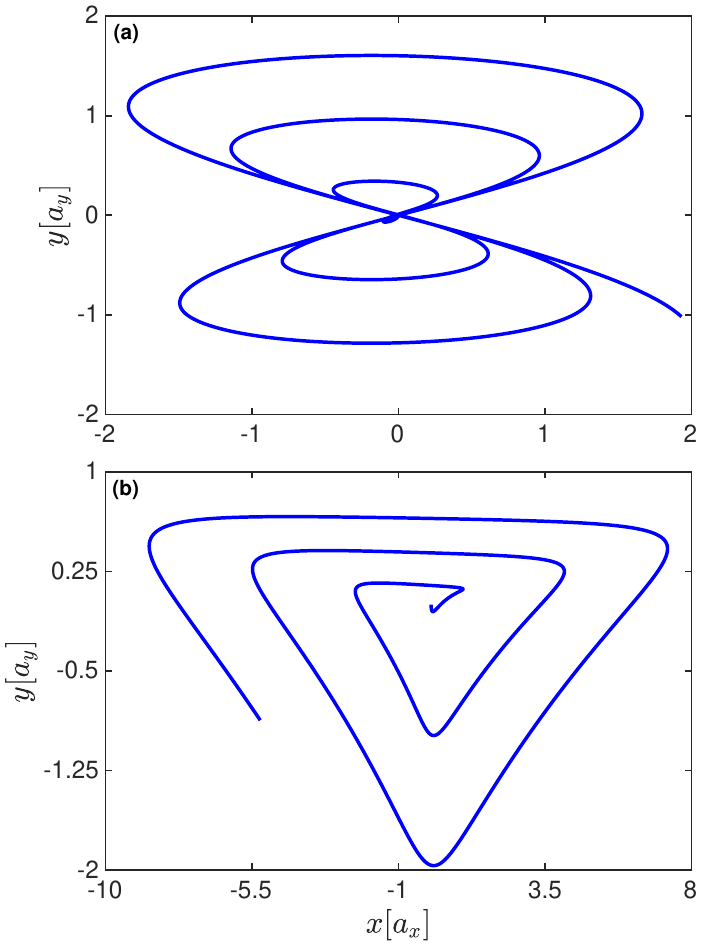,width=\columnwidth}}
\caption{Lissajous-like figure of Bloch particle for monolayer phosphorene:
(a) for $k_yd_2=\frac{\pi}{2}$, (b) for $k_yd_2=\frac{\pi}{4}$.
Other parameters used in the numerical
simulations are the same as used for Fig.~\ref{Figure2}.}
\label{Figure9}
\end{figure}
As a consequence, the Bloch velocity is affected significantly by applying
an in-plane electric field, which oscillates with oscillation frequency
$\omega_x=\frac{eE_xd_1}{\hbar}$ showing its periodic character,
i.e., $v_j\left(t+T\right)=v_j\left(t\right)$ with
$T=\frac{2\pi}{\omega_x}$ being the time period of the motion.
It is shown that $v_y\left(t\right)$ is
modified strongly even if the electric field is applied in the
$x$-direction because the energy dispersion couples the $x$-
and $y$-components of the crystal momentum.
In addition, it is obvious that for
increasing value of $k_y$, the wave packet begins
to wind the Brillouin zone in two different directions with angular
frequency $\omega_x$.
In Fig.~\ref{Figure7}, we show the Bloch velocity $v_x$ as a function
of time with oscillation $\omega_x$ under the influence of an in-plane
electric field applied in the $x$-direction. Fig.~\ref{Figure7} (a)
reveals that the amplitude and phase of oscillations are modified
considerably by changing the initial crystal momentum $k_y$, see the
blue solid, black dashed, and green dash-dotted curves in panel (a).
Similar features of $v_y$ can be seen in Fig.~\ref{Figure7} (b).
In addition, comparison of panels (a) and (b) reveals different
dynamical behavior of the Bloch electron in the $x$- and $y$-directions.
In particular, the $x$-component of the Bloch velocity oscillates with
large frequency compared to the $y$-component. Moreover, the $y$-component
of the Bloch velocity vanishes for $k_y=\pi/d_2$.
For further analysis, the real space trajectories of the Bloch
dynamics as a function of time are shown in Fig.~\ref{Figure8}.
This figure also reveals oscillatory behaviour of the Bloch dynamics
in real space, depending on the initial crystal momentum $k_y$
as is obvious from comparison of the blue solid, black dashed,
and green dash-dotted curves in panels (a) and (b), where
the change in oscillation frequency and amplitude is obvious.
Interestingly, the amplitude of oscillation increases with the increase in time.
Moreover, we plot the
real-space trajectories of the Bloch oscillations in Fig.~\ref{Figure9}
for two different values of the initial $k_y$ momentum which
exhibits Lissajous-like oscillations. It is obvious that with
increasing value of $k_y$, wave packet starts
to wind Brillouin zone in two different directions
with angular frequency $\omega_x$. Comparison of panels (a) and (b)
reveals strong dependence of the dynamics on the initial crystal momentum $k_y$.
\subsection{Bloch dynamics in an in-plane electric field along $y$-axis}
\label{Sec:Hamiltonian}
Here we consider the case when the electric field is applied
in the $y$-direction, i.e., $E=E_y$, hence $E_x=0$.
In this case,
the semiclassical dynamical equation shows that
$k_x(t)=k_x=\textrm{constant}$ and $k_y(t)=k_y(0)+\frac{eE_y}{\hbar} t$.
Hence, the Bloch velocity is affected significantly by applying
an in-plane electric field in the $y$-direction, which oscillates with oscillation frequency
$\omega_y=\frac{eE_yd_2}{\hbar}$ showing its periodic character,
i.e., $v_j\left(t+T\right)=v_j\left(t\right)$ with
$T=\frac{2\pi}{\omega_y}$ being the time period of the motion.
\begin{figure}[!ht]
\centerline{\psfig{figure=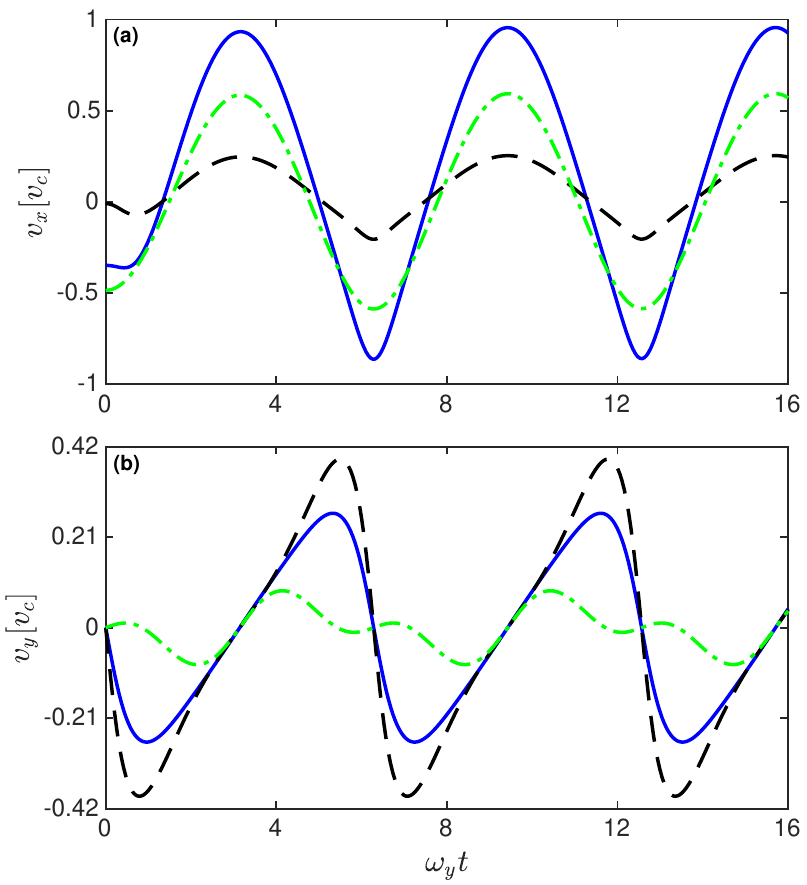,width=\columnwidth}}
\caption{Bloch velocity as a function of time for monolayer
phosphorene with $\omega_y=\frac{eE_yd_2}{\hbar}$.
Panel (a) shows the Berry velocity in the $x$-direction, whereas (b) in the $y$-direction.
In each panel, the blue solid curve is used for $k_xd_1=\frac{\pi}{2}$, black dashed curve for $k_xd_1=\frac{\pi}{4}$, and green dash-dotted curve for $k_xd_1=\pi$.
Other parameters used in the numerical
simulations are the same as used for Fig.~\ref{Figure2}.}
\label{Figure10}
\end{figure}
\begin{figure}[!ht]
\centerline{\psfig{figure=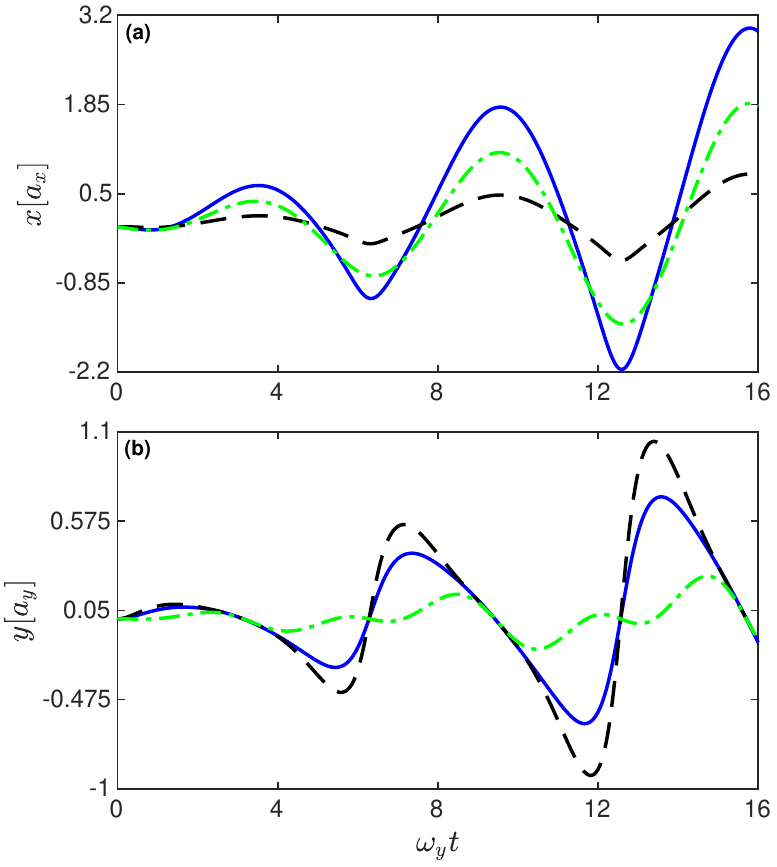,width=\columnwidth}}
\caption{Real space trajectories of Bloch particle as a function
of time for monolayer phosphorene for: (a) $x$-component, (b)
$y$-component, where in each panel, the blue solid curve is used
for $k_xd_1=\frac{\pi}{2}$, black dashed curve for $k_xd_1=\frac{\pi}{4}$,
and green dash-dotted curve for $k_xd_1=\pi$.
Other parameters used in the numerical
simulations are the same as used for Fig.~\ref{Figure2}.}
\label{Figure11}
\end{figure}
\begin{figure}[!ht]
\centerline{\psfig{figure=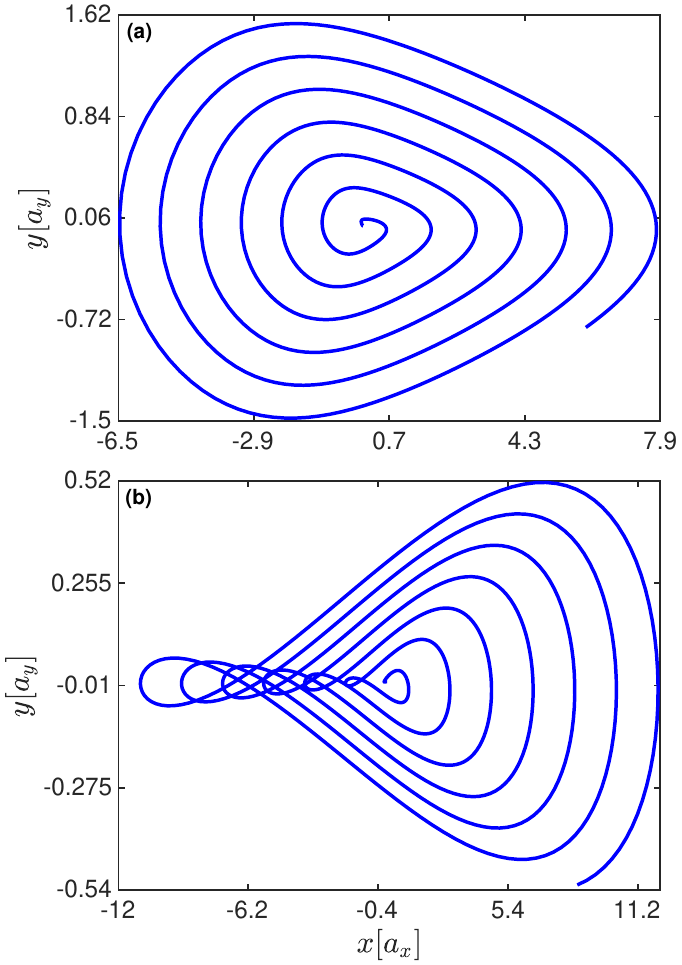,width=\columnwidth}}
\caption{Lissajous-like figure of Bloch particle for monolayer phosphorene:
(a) for $k_xd_1=\frac{\pi}{2}$, (b) for $k_xd_1=\frac{\pi}{4}$.
Other parameters used in the numerical
simulations are the same as used for Fig.~\ref{Figure2}.}
\label{Figure12}
\end{figure}
In Fig.~\ref{Figure10}, we show the Bloch velocity as
a function of time in monolayer phosphorene with broken inversion
symmetry in an in-plane electric field $\textbf{E}$
applied in the $y$-direction, using $k_xd_1=\frac{\pi}{2}$, see blue solid
curves, $k_xd_1=\frac{\pi}{4}$, see black dashed curves, and $k_xd_1=\pi$,
green dash-dotted curves in both panels (a) and (b). Comparison of panels
(a) and (b) reveals that the wave packet undergoes pronounced oscillatory
motion in monolayer phosphorene under the influence of an in-plane
electric field. In addition, Fig.~\ref{Figure10} (a)
shows that the wave packet exhibits finite Bloch velocity in the
$x$-direction even when the electric field is applied in the $y$-direction.
Interestingly, comparison of panels (a) and (b) reveals that $v_x\left(t\right)$
and $v_y\left(t\right)$ perform out of phase oscillations with different amplitudes.
Moreover, comparison of Figs.~\ref{Figure7} and~\ref{Figure10} shows
that the Bloch velocity exhibits different dynamical behavior under the
influence of applied in-plane electric field in the $x$- and $y$-directions.
To realize the real space dynamics,
we show the real space trajectories in Fig.~\ref{Figure11}
using the same set of parameters as used for Fig.~\ref{Figure10}.
This figure reveals pronounced oscillatory behavior of the
system dynamics. Comparison of the blue solid, black
dashed, and green dash-dotted curves in both panels (a) and (b)
reveals that the Bloch dynamics is significantly affected by the
initial momentum $k_x$. Likewise, comparison of panels (a) and (b)
shows that the $x$- and $y$-components of the Bloch dynamics
exhibits different dynamical behaviour.
For further understanding, we plot the
real-space trajectories of the Bloch oscillations in Fig.~\ref{Figure12}
for two different values of the initial $k_x$ momentum which
exhibits Lissajous-like oscillations. Comparison of panels (a) and (b)
reveals the strong dependence of Bloch dynamics on the initial momentum $k_x$.
Moreover, comparison of Figs.~\ref{Figure9} and~\ref{Figure12} shows
the difference in dynamical behavior of Bloch dynamics under the
influence of applied in-plane electric field in the $x$- and $y$-directions.
\subsection{Effect of spin-orbit interaction on Bloch dynamics}\label{SOIDC}
In this section, the effect
of spin-orbit interaction (SOI) on the Bloch dynamics in
monolayer phosphorene with broken inversion symmetry is investigated.
This study is expected to be useful in understanding the spin-dependent
electronic properties that may pave the way for
potential applications of phosphorene in spintronic devices.
Interesting effects are induced by the spin-orbit interaction in phosphorene~\cite{Kurpas-PRB.94:155423,Farzaneh-PRB.100:245429,Popovic-PRB.92:035135}.
The details of spin-orbit interaction in phosphorene can be
found in~\cite{Yar-JPCM.35:165701,Sultana-JPCS.176:111257}. Here we focus merely
on its impact on Bloch oscillations.
In this paper, the effects of spin-orbit interaction
are incorporated considering the intrinsic spin–orbit
coupling within the framework of Kane–Mele model which
takes into account appropriately the effects of spin up
and spin down states as used in phosphorene~\cite{Rezania-EPJP.137:18,Sultana-JPCS.176:111257,Yar-JPCM.35:165701}, borophene~\cite{Yar-PLA.429:127916}, lattice system~\cite{Haldane-PRL.61:2015}, graphene~\cite{Kane-PRL.95:226801}, and, silicene~\cite{Vargiamidis-JPCM.26:345303}.
\begin{figure}[!ht]
\centerline{\psfig{figure=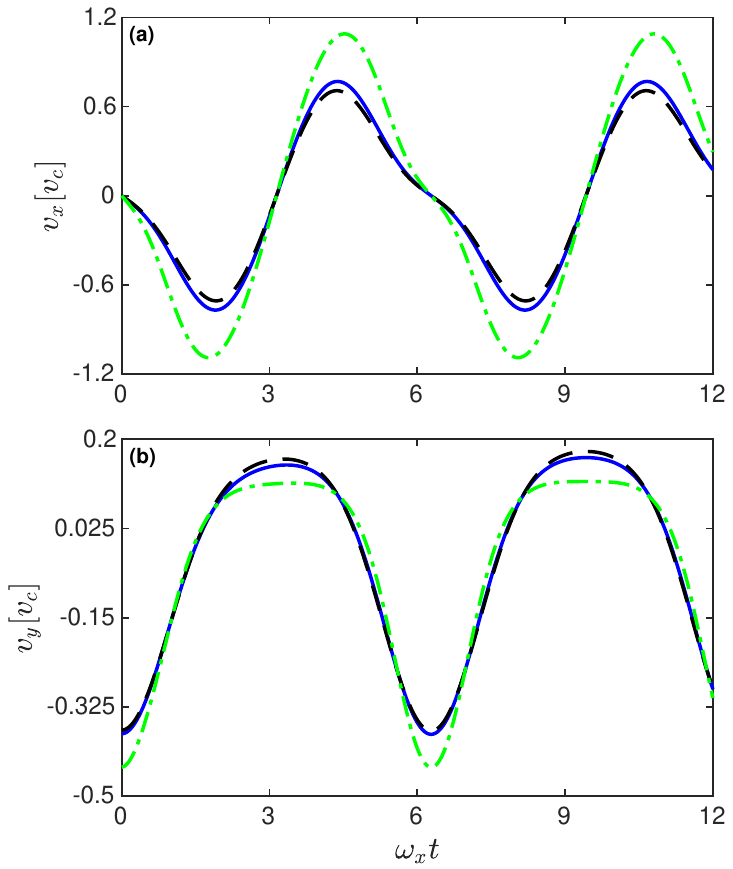,width=\columnwidth}}
\caption{Bloch velocity as a function of time for monolayer phosphorene,
illustrating the effect of spin-orbit interaction.
Panel (a) shows the Berry velocity in the $x$-direction, whereas (b) in the $y$-direction.
In each panel, the blue solid curve is used for $k_yd_2=\frac{\pi}{2}$, black dashed curve for $k_yd_2=\frac{\pi}{4}$, and green dash-dotted curve for $k_yd_2=\pi$.
The parameters used are:
$\Delta=\delta,\ \Delta_{SO}=0.6\delta, \ \Delta_z=2\delta$ and
other parameters used in the numerical
simulations are the same as used for Fig.~\ref{Figure2}.}
\label{Figure13}
\end{figure}
\begin{figure}[!ht]
\centerline{\psfig{figure=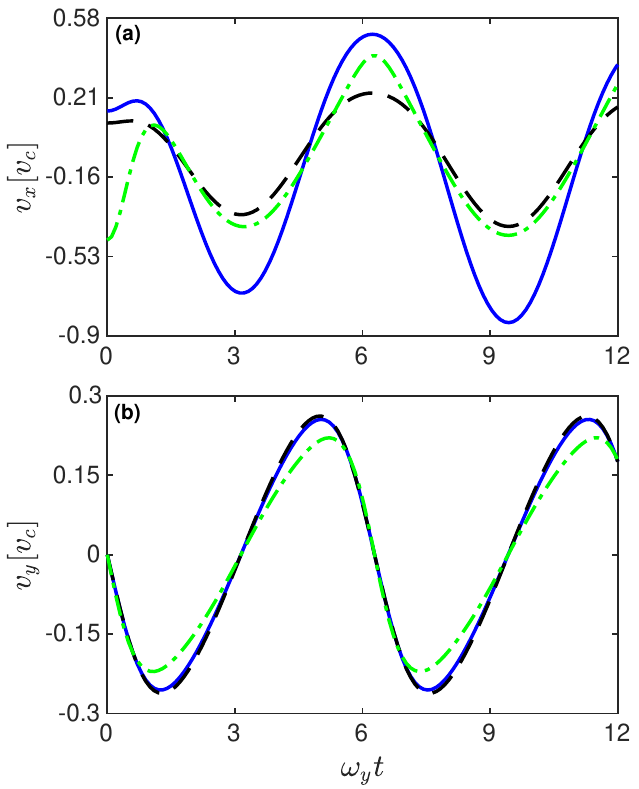,width=\columnwidth}}
\caption{Bloch velocity as a function of time for monolayer phosphorene,
illustrating the effect of spin-orbit interaction.
Panel (a) shows the Berry velocity in the $x$-direction, whereas (b) in the $y$-direction.
In each panel, the blue solid curve is used for $k_xd_1=\frac{\pi}{2}$, black dashed curve for $k_xd_1=\frac{\pi}{4}$, and green dash-dotted curve for $k_xd_1=\pi$. The parameters used in the numerical
simulations are the same as used for Fig.~\ref{Figure12}.}
\label{Figure14}
\end{figure}
The Hamiltonian of monolayer phosphorene with broken inversion
symmetry under the influence of
intrinsic spin-orbit interaction can be described as
\begin{align}\label{eq:HamiltonianSOI}
\mathcal{H}\left(\textbf{k}\right) &=\mathcal{H}_0\left(\textbf{k}\right)
+\mathcal{H}_{SOI}\left(\textbf{k}\right),
\end{align}%
where $\mathcal{H}_0\left(\textbf{k}\right)$ is given in Eq.~\eqref{eq:Hamiltonian2by2}, whereas $\mathcal{H}_{SOI}\left(\textbf{k}\right)=\Delta_z\sigma_z-s_z\Delta_{SOI}\sigma_z$
characterizes the Kane-Mele Hamiltonian, denoting the
intrinsic spin-orbit interaction (SOI) and induces the
SOI gap, $\Delta_{SOI}$, in the energy spectrum of the system.
The factor, $\Delta_z=lE_z$ with the length scale $l=2.26\textrm{\AA}$,
takes into account the
effects of electric field $E_z$ applied perpendicular to the sample.
Likewise, $s_z=\pm $ stands for the spin direction such that $s_z=+$
represents spin up and $s_z=-$ characterizes the spin down state.
The Hamiltonian in Eq.~\eqref{eq:HamiltonianSOI} can be diagonalized
using the standard diagonalization method. Using the obtained eigenenergies,
one can readily evaluate the velocities of the Bloch electron.
In Fig.~\ref{Figure13}, we show the Bloch velocity as a function of time using
$\Delta=\delta,\ \Delta_{SOI}=0.6\delta, \ \Delta_z=2\delta$,
where panel (a) represents the $x$-component and
(b) the $y$-component under the influence of an
in-plane electric field in the $x$-direction. In each panel,
the green dash-dotted curve shows the
Bloch dynamics without spin-orbit coupling, the blue solid curve
for spin up, whereas the black dashed curve for spin down states.
Comparison of the blue solid, black dashed,
and green dash-dotted curves in both panels (a) and (b) shows that
the spin-orbit interaction remarkably changes the Bloch oscillations,
depending on the strength of interaction. Moreover, comparison
of panels (a) and (b) reveals that the effect of SOI is more pronounced
on the $x$-component of the Bloch velocity compared to the $y$-component.
In addition, comparison of the blue solid and black dashed curves shows that
the response of the spin up and spin states are different.
In Fig.~\ref{Figure14}, we show the effect of spin-orbit coupling
on the velocity of Bloch electron in monolayer phosphorene with
broken inversion symmetry when the in-plane electric is applied
in the $y$-direction. Comparison of the blue solid, black dashed,
and green dash-dotted curves in both panels (a) and (b) shows that
the spin-orbit interaction changes the Bloch oscillations considerably,
depending on the strength of interaction. Moreover, comparison
of panels (a) and (b) reveals that the effect of SOI is more pronounced
on the $x$-component of the Bloch velocity compared to the $y$-component.
Further comparison of the blue solid and black dashed curves shows that
the response of the spin up and spin states are different.
Furthermore, comparison of Figs.~\ref{Figure13} and~\ref{Figure14} shows that the SOI
affects differently when the in-plane electric field is applied in the $x$- and $y$-directions.
\subsection{Confined-deconfined state transition}
\label{Sec:Hamiltonian}
In this section, we study the effect of in-plane electric
and transverse magnetic fields on the Bloch dynamics in monolayer
phosphorene which essentially leads to a transition from confined
to deconfined states and vice versa that strongly depend on
the relative strength of the fields.
\begin{figure}[!ht]
\centerline{\psfig{figure=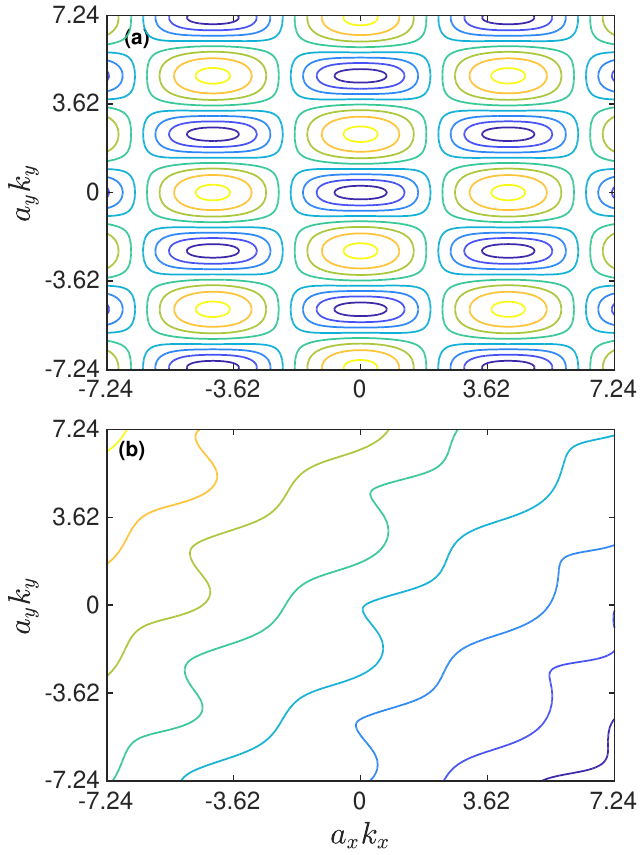,width=\columnwidth}}
\caption{Confined-to-deconfined transition by analyzing the
trajectories of $H\left(k_x,k_y\right)$:
(a) $eE = 0.1F_0$, $B = 6 T$; (b) $eE = 6F_0$, $B = 0.1 T$,
which are contours in momentum space, depending on the strength
of electric and magnetic. Other parameters remain the same
as used for Fig.~\ref{Figure2}.}
\label{Figure15}
\end{figure}
In this case, the wave packet
dynamics in conduction band is determined using the semiclassical
dynamical equation
\begin{align}\label{eq:kEB}
\hbar\dot{\textbf{k}}=e\textbf{E}+e\textbf{v}\times\textbf{B},
\end{align}
where $\textbf{E}$ is the applied electric field and $\textbf{B}$ is the magnetic field.
Solving Eqs.~\eqref{eq:SCv},~\eqref{eq:vx},~\eqref{eq:vy}, and~\eqref{eq:kEB},
we can study the Bloch dynamics in a monolayer phosphorene with broken inversion symmetry.
The position $\textbf{r}(t) = \int^t_0 \textbf{v}(t')dt'$ can be determined
by integrating the equation of motion:
\begin{align}\label{eq:EOM}
\hbar\left[\textbf{k}(t)-\textbf{k}(0)\right]=e\textbf{E}t+e\textbf{r}(t)\times\textbf{B}.
\end{align}
In the confined (B-dominated) regime, the drift velocity
$\textbf{v}_d = \textbf{r}(t)/t|_{nT}$ is given by
\begin{align}\label{eq:vd}
\textbf{v}_d=\textbf{E}\times\textbf{B}/B^2.
\end{align}
In the transition to deconfined (E-dominated) regime, the
drift velocity abruptly drops to zero.
Interesting dynamics appears in an applied transverse magnetic
field, where dynamical phase transition to one-frequency oscillation occurs.
As a consequence, the system exhibits complex dynamics at the transition.
It is shown that under the influence of in-plane electric and transverse
magnetic fields, two distinct types
of cyclotron orbits are formed depending on the relative strength
of $\textbf{E}$ and $\textbf{B}$:
(i) when magnetic field dominates the in-plane electric field, confined
orbits are formed which reside within the Brillouin zone
and characterized by one Bloch frequency,
(ii) however, de-confined orbits are generated when E-field
dominates B-field which extend over infinitely many
Brillouin zones and are described by two or more frequencies.
It is illustrated that confinement in $k$-space means
deconfinement in $r$-space, and vice versa.
Here the equations of motion can be determined in terms
of a Hamiltonian function as~\cite{Yar-PLA.478:128899}
\begin{align}\label{eq:EOM}
\dot{k}_x=\frac{\partial H\left(k_x,k_y\right)}{\partial k_y},\quad
\dot{k}_y=-\frac{\partial H\left(k_x,k_y\right)}{\partial k_x},
\end{align}
where the Hamiltonian function is defined as
\begin{align}\label{eq:HamilFunc}
H\left(k_x,k_y\right)=\frac{eB}{\hbar^2}E\left(\textbf{k}\right)
+\frac{e}{\hbar}\left|\textbf{E}\times \textbf{k}\right|,
\end{align}
where $E\left(\textbf{k}\right)$ denotes the energy dispersion and
$\textbf{E}$ characterizes the applied electric field.
The trajectories of wave packet appear as contours of
$H\left(k_x,k_y\right)$ in momentum space.
The effects of electric and magnetic fields on the Bloch dynamics
are incorporated appropriately using Eq.~\eqref{eq:HamilFunc}.
Note that the trajectories are confined orbits in the Brillouin zone
with single frequency in the regime, $E<vB$, whereas de-confined orbits
are formed which are extended over infinitely many
Brillouin zones with two or more frequencies when $E>vB$.
To highlight this effect, the contours of the Hamiltonian function
in Eq.~\eqref{eq:HamilFunc} are plotted as a function of
crystal momenta, $k_x$ and $k_y$, in Fig.~\ref{Figure15}, illustrating the confinement and deconfinement
of orbits which depend on the relative strength of the electric and magnetic fields.
This figure shows that the orbits are confined in the regime $E<vB$,
see Fig.~\ref{Figure15}(a), however the orbits exhibit
de-confined behaviour when the strength of electric field is greater
than the magnetic field, i.e., $E>vB$, see Fig.~\ref{Figure15}(b).
\section{Conclusions}\label{Sec:Conc}
In summary, we have studied Bloch dynamics in monolayer
phosphorene with broken inversion symmetry within the
framework of semiclassical theory.
We have shown that the Bloch velocity of a
wave packet exhibits pronounced oscillations in both
real and momentum spaces, called Bloch oscillations.
It has been found that an applied in-plane electric
field modifies significantly the Bloch oscillations in the
system, depending on its magnitude and direction.
Dynamical transition is driven by an applied magnetic field,
leading to a complex dynamics at the transition point.
In the presence of both external
in-plane electric and transverse magnetic fields,
the system undergoes a dynamical transition from confined to
de-confined state and vice versa, tuned by
the relative strength of the applied fields which was also
observed in a moir\'{e} flat band system~\cite{Yar-PLA.478:128899}.
In this case, two distinct types
of cyclotron orbits are formed, depending on the relative strength
of $\textbf{E}$ and $\textbf{B}$:
(i) when magnetic field dominates the in-plane electric field, confined
orbits are formed which reside within the Brillouin zone
and characterized by a single Bloch frequency,
(ii) however, de-confined orbits are generated when E-field
dominates B-field which extend over infinitely many
Brillouin zones and are described by two or
more frequencies. The equations of motion can be
derived by defining a Hamiltonian function $H\left(k_x,k_y\right)$
with trajectories in the form of contours in momentum space.
It has been shown that the confinement of orbits depends
on the relative strength of electric and magnetic fields such that the
orbits are confined when the strength of magnetic field
is greater than the electric field, i.e., $vB>E$, which
however become de-confined for $vB<E$.
It is illustrated that the Bloch dynamics in monolayer phosphorene
with broken inversion symmetry presents a dynamical scenario that differs from the
Bloch oscillations in moir\'{e} flat band system~\cite{Yar-PLA.478:128899}. For instance,
in the present study, we have focussed on the investigation of Bloch
velocity composed of Berry and group velocities, whereas in the latter
system we have studied the group velocity only
with focus on the effect of twist angle with preserved inversion symmetry of the system.
Due to the difference in models, the results of the two systems are very different.
However, in both systems we have studied the Bloch oscillations under the
influence of external fields such as in-plane electric and
transverse magnetic fields, where in both systems the wave
packets exhibit pronounced Bloch oscillations and the system undergoes a dynamical transition.
The experimental measurement of Bloch oscillations
in monolayer phosphorene with broken inversion symmetry
is expected to be possible using the techniques developed for observing
oscillations on the surface of black phosphorus using a gate
electric field~\cite{Li-NN.10:608},
transport measurements of phosphorene-hexagonal BN (hBN)
heterostructures with one-dimensional edge contacts~\cite{Gillgren-TDM.2:011001}, and
time-resolved band gap emission spectroscopy~\cite{Li-OE.26:23844}.
\section*{Acknowledgments}
A. Yar acknowledges the support of Higher Education Commission (HEC),
Pakistan under National Research Program for Universities NRPU Project No. 11459.
\section*{Data availability statement}
Data sharing is not applicable to this article, as it describes
entirely theoretical research work.


\end{document}